\documentclass{article}

\usepackage{arxiv}

\usepackage{amsmath}
\usepackage{comment}
\usepackage{bigints}
\usepackage{amssymb}
\usepackage[utf8]{inputenc} 
\usepackage{graphicx}
\usepackage[font={small}]{caption}
\usepackage{subcaption}
\usepackage[T1]{fontenc}    
\usepackage{url}            
\usepackage{booktabs}       
\usepackage{algorithm}
\usepackage{algorithmic}

\DeclareMathOperator*{\argmin}{argmin}
\usepackage{algorithm}
\usepackage{algorithmic}
\usepackage{amsfonts}       
\usepackage{nicefrac}       
\usepackage{microtype}      
\usepackage{lipsum}
\usepackage{hyperref}
\usepackage{mathtools} 
\usepackage{empheq}

\hypersetup{
    colorlinks=true,
    linkcolor=blue,
    filecolor=magenta,      
    urlcolor=cyan,
}

\title{Cluster-based Regression using Variational Inference and Applications in Financial Forecasting}
\author{Udai G. Nagpal \thanks{Udai Nagpal is an Associate in Asset Management in Goldman Sachs \& Co. The opinions expressed here are those of the author and do not represent those of his employer Goldman Sachs \& Co. and its affiliates.} \\
\texttt{ugnagpal@gmail.com} 
 \And
Krishan M. Nagpal \thanks{Krishan M. Nagpal is a Managing Director in Corporate Risk in Wells Fargo \& Co. The opinions expressed here are those of the author and do not represent those of his employer Wells Fargo \& Co. and its affiliates.}\\
\texttt{krishan$\_$nagpal@yahoo.com}
}
\begin{document}
\maketitle

\begin{abstract}
This paper describes an approach to simultaneously identify clusters and estimate cluster-specific regression parameters from the given data. Such an approach can be useful in learning the relationship between input and output when the regression parameters for estimating output are different in different regions of the input space. Variational Inference (VI), a machine learning approach to obtain posterior probability densities using optimization techniques, is used to identify clusters of explanatory variables and regression parameters for each cluster. From these results, one can obtain both the expected value and the full distribution of predicted output. Other advantages of the proposed approach include the elegant theoretical solution and clear interpretability of results. The proposed approach is well-suited for financial forecasting where markets have different regimes (or clusters) with different patterns and correlations of market changes in each regime. In financial applications, knowledge about such clusters can provide useful insights about portfolio performance and identify the relative importance of variables in different market regimes. An illustrative example of predicting one-day S\&P change is considered to illustrate the approach and compare the performance of the proposed approach with standard regression without clusters. Due to the broad applicability of the problem, its elegant theoretical solution, and the computational efficiency of the proposed algorithm, the approach may be useful in a number of areas extending beyond the financial domain.


\end{abstract}

\section{Introduction}

In many regression applications with cluster-specific patterns between inputs and outputs, output predictions can be improved by making regression parameters dependent on the input region or cluster. The difficulty lies in optimally identifying clusters so that regression based on those clusters provides the best possible estimate of the output. While the proposed approach may be applicable in many areas, in the subsequent sections we will focus on financial markets to illustrate the approach. Cluster based prediction approaches are well-suited for financial applications as markets exhibit regimes where in each regime market changes and correlations are similar. For example, market participants view market conditions as "rates increasing" vs. "rates decreasing" environments, or as "risk on" vs. "risk off" environments, where in each of such environments market changes are expected to follow a regime-specific pattern.

Variational inference (VI) is generally used to obtain analytical approximations of the posterior probability density for graphical models where the observed data is dependent on unobserved latent variables. VI has been used in a wide range of applications to get approximate Bayesian posterior distributions conditioned on data (see for example (\cite{Bishop} and \cite{BleiStats}). Compared to other approaches based on sampling such as the Metropolis-Hastings algorithm (\cite{Gelfand}), VI can be more computationally efficient for large datasets or complex distributions. The concept of latent variables in creating posterior distributions in VI has a natural analog in financial data as market participants think of market conditions as different "regimes" where in each regime markets behave in a particular manner. In VI or other latent variable approaches, these market regimes such as "risk on" or "risk off" are not user-specified but rather identified through analysis of the data.

In the proposed framework it is assumed that the underlying data is a mixture model of Gaussian clusters with a locally linear mapping where the unknown linear regression vector for each cluster is drawn from another Gaussian distribution. Similar mixture models have been considered in \cite{dele} and for dynamical systems in \cite{pineda} and in both cases parameter estimation is based on Expectation-Maximization. Variational approaches for mixtures of linear mixed models have also been considered \cite{tan}. Utilizing a similar VI framework, performance of the proposed approach can potentially be enhanced by replacing linear relationship between inputs and the output by a cluster-specific neural network to incorporate nonlinearities in the input-output relationship. However, obtaining posterior distribution of neural network parameters and the model can be challenging (see for example \cite{shridhar}). While replacing linear mapping with neural networks may provide more accurate prediction, there are some important advantages of the proposed approach over neural variational inference such as a) theoretically elegant solution with closed form expression for distribution of parameters and the output, b) fast calibration for obtaining model parameters, and c) greater transparency and interpretability of the model which is desired by most practitioners (interpretability is also discussed in the example section). As in traditional VI approaches, it is assumed that the datasets are not too large for memory requirements. For calibration of graphical models with very large datasets, approaches such as stochastic variational inference may be more useful where parameter updates are based on stochastic optimization (see for example \cite{hoffman}).



Due to difficulty in forecasting financial market data, various types of approaches ranging from fundamental and econometric to technical charts and machine leaning approaches have been proposed with varying levels of success (see for example \cite{shen}). Some of the early approaches for forecasting equity changes were based on regression and time series prediction models such as ARMA. More complex modeling approaches for dynamical systems such as Hidden Markov Models have also been proposed (see for example \cite{hassan}). Neural networks of various forms have also been proposed (see for example \cite{Enke}) and are generally good at identifying non-linear relationships. Ensemble of multiple decision trees can be used to synthesize forecasts from different indicators as in \cite{Khaidem}. Support Vector Machines have shown strong performance in financial forecasting (see for example \cite{Kim}). 

This paper is organized as follows. The next section contains a brief introduction to variational inference and the mean-field family of distributions. The underlying assumptions and modeling framework are described in Section 3. Section 4 contains illustrative examples to demonstrate the utility of the proposed approach in both normal and volatile market conditions. Section 5 concludes with a summary and the Appendix contains the derivation of the variational parameter estimates.

\section{Brief Introduction to Variational Inference}

In Bayesian statistics the goal is to infer the posterior distribution of unknown quantities using observations. For some complex problems it is assumed that observations, denoted by $x=\{x_1,\cdots,x_N\}$ in this section, are linked to latent variables $z=\{z_1,\cdots,z_K\}$ which are drawn from prior distribution $p(z)$. The likelihood of observation $x$ depends on $z$ through the distribution $p(x|z)$. In Bayesian learning and estimation problems the goal is to learn the posterior distribution $p(z|x)$ of latent variables conditioned on the data.

VI is a widely-used method in Bayesian machine learning for approximating posterior distributions. In VI a suitable family of distributions $\mathcal(F)$ is chosen which is complex enough to capture the attributes of the data and yet simple enough to be computationally tractable. Then, the best approximate posterior distribution is obtained from an optimization to minimize Kullback-Leibler (KL) divergence with the exact posterior (see for example (\cite{Bishop} and \cite{BleiStats}): 
\begin{equation}
\label{q*def}
    q^*(z)=\argmin_{q(z) \in \mathcal(F)} \; KL \left( q(z) || p(z|x) \right)
\end{equation}
where the KL divergence, a measure of information-theoretic distance between two distributions, is obtained as follows where $\mathbb{E}_q$ denotes expectation with respect to $q(z)$:
\begin{align}
    KL \left( q(z) || p(z|x) \right) & :=  \mathbb{E}_q \left[\text{log} \ q(z)\right] - \mathbb{E}_q \left[ \text{log} \ p(z|x)\right] \nonumber \\
    \label{KL}
    & = \mathbb{E}_q \left[ \text{log} \ q(z)\right] - \mathbb{E}_q \left[ \text{log} \ p(z,x)\right] + \text{log} \ p(x)
\end{align}
Since log $p(x)$ does not depend on the distribution $q(z)$, the optimal distribution $q^*(z)$ that minimizes KL divergence in (\ref{q*def}) also maximizes the Evidence Lower Bound (ELBO) defined as follows:
\begin{equation}
\label{elbo}
    \text{ELBO}(q):=\mathbb{E}_q \left[ \text{log} \ p(z,x)\right] -\mathbb{E}_q \left[ \text{log} \ q(z)\right]
\end{equation}

Here we will work with the mean-field variational family of distributions $q(z)$ in which one assumes that the latent variables are mutually independent and probability density of each latent variable is governed by distinct factors (see \cite{Bishop} and \cite{BleiStats} for more details). This assumption greatly simplifies the computation of the optimal parameters of the variational distribution $q(z)$. Due to the assumption of mutual independence, if there are $K$ latent variables $z=\{z_1, \hdots, z_K\}$,  the joint density function is of the form
\[
q(z)= \prod_{k=1}^K q_k(z_k)
\]
where $q_k(z_k)$ is the density of the k'th latent variable $z_k$ and each $q_k$ has its own set of parameters. For the mean-field variational family of distributions, the most commonly used approach for obtaining parameters that maximize the ELBO (defined in (\ref{elbo})) is coordinate ascent variational inference (CAVI). In this approach, which has similarities to Gibbs sampling, one optimizes parameters for each latent variable one at a time while holding others fixed, resulting in a monotonic increase in the ELBO (\cite{Bishop},\cite{BleiStats}). 

It can be shown (\cite{Bishop}) that for the mean-field variational family, $q_k^*(z_k)$, the optimal distribution for $q_k(z_k)$ that maximizes the ELBO (\ref{elbo}), satisfies the following: 
\[
    q_k^*(z_k)\propto \text{exp} \left\{ \mathbb{E}_{-k} \left[ \text{log} \ p(z_k \mid z_{-k}, x)\right] \right\}
\]

where $z_{-k}$ denotes all $z_i$ other than $z_k$ and $\mathbb{E}_{-k}$ represents expectation with respect to all $z_i$ other than $z_k$. Thus the optimal variational distribution of latent variable $z_k$ is proportional to the exponentiated expected log of the posterior conditional distribution of $z_k$ given all other latent variables and all the observations. Since $p(z_{-k}, x)$ does not depend on $z_k$ (due to independence of latent variables $z_i$), one can write the above equivalently as
\begin{equation}
\label{meanfield}
    q_k^*(z_k)\propto \text{exp} \left\{ \mathbb{E}_{-k} \left[ \text{log} \ p(z_k , z_{-k}, x)\right] \right\}
\end{equation}
For example, if there are three latent variables $z_1,z_2$ and $z_3$, the above implies
\[
q_1^*(z_1)\propto \text{exp} \left\{ \mathbb{E}_{z_2,z_3} \left[ \text{log} \ p(z_1, z_2, z_3, x)\right] \right\}
\]
The proposed approach is based on the mean-field family and thus (\ref{meanfield}) will be instrumental in obtaining the parameters of the posterior distribution.

\section{Underlying Model and VI Calibration}

It is assumed that the available information known at time $t$ is $x_t \in \mathbb{R}^n$ and this is used to predict some signal $y_{t+1} \in \mathbb{R}$ that is not known until time $t+1$. In financial applications, vector $x_t$ could be a function of observed market data (such as S$\&$P index, swap rates or equity implied volatilities) and their recent trend, trading volumes etc. The optimal choice of components of $x_t$ could be based on choosing the combination that has the best out-of-sample performance using the proposed algorithm. The goal is to predict signal $y_{t+1}$ and its distribution at time $t+1$ from $x_t$, which comprises information available at time $t$. As an illustrative example, let us assume that our goal is to predict S$\&$P change for date $t+1$ based on available information up to date $t$. Here $y_{t+1}$ will be S$\&$P index change from day $t$ to day $t+1$. The input $x_t$, which is based only on information up to time $t$, could be composed of the prior day change in S$\&$P index, trading volume deviation from average, implied volatility in recent days, changes in interest rates, etc. 

We will assume that there are K clusters where each cluster can be thought of as a different market regime. For any time $t$, $x_t$ belongs to one of the K clusters. For each $t$, $c_t \in \mathbb{R}^K$ will denote the indicator vector that describes which cluster $x_t$ belongs to. The indicator vector $c_t$ has $K-1$ elements equal to $0$ and one element equal to $1$ that corresponds to the cluster assignment of $x_t$. For example, if $x_t$ is in the second cluster then 
\[
c_t= \begin{bmatrix}
    0  \\ 1 \\ 0 \\ \vdots \end{bmatrix}
\]
We will denote $c_{t(k)}$ as the $k'th$ element of $c_t$ and as noted above $c_{t(k)}=1$ for only one $k \in \{1, \cdots, K \}$ and is zero for all other $k$. For example if the vector $c_t$ is as above, $c_{t(2)}=1$ and $c_{t(k)}=0$ for all $k \neq 2$. We will assume that the mean and variance of cluster means do not change over time, and we will denote the likelihood that the state $x_t$ is in cluster $k$ as $\pi_k$ (the sum of $\pi_k$ over $K$ has to be one as at every time $x_t$ is in one of the $K$ clusters). The assumption that cluster probability $\pi_k$ does not depend on time $t$ does not imply that the underlying time series must be stationary. For non-stationary data such as equity indices, $x_t$ would be obtained by transforming non-stationary time series to stationary signals. For example, for non-stationary equity time series, $x_t$ could be composed of drift terms (such as change over $n$ days) which are stationary. In other words, the dynamical aspects of the markets (such as momentum and/or mean reversion) here would be captured by appropriate transformations of market data such as changes over $n$ days in the S$\&$P index or interest rates. 

The mean of $x_t$ for cluster $k$, which we denote by $\mu_k$, is not known. It is assumed that for each cluster $k$, $\mu_k$ is normally distributed with mean $\mu_{k0}$ and variance $R_{k0}$. Within each cluster $k$, $x_t$ is normally distributed around the cluster mean $\mu_k$ with variance $M$, which is the same for all clusters. It is assumed that explanatory variables $x_t$ have been suitably adjusted for mean and drift terms so that the cluster mean $\mu_k$ does not depend on time $t$.

It is assumed that for each of the $K$ clusters there are regression vectors $\beta_k$ that linearly map $x_t$ to the output of interest $y_{t+1}$. More specifically:

{\centering If $x_t$ is in cluster $k$ then $y_{t+1}=\beta_k'x_t+v_{t+1}$ where noise $\{v_{t+1}\} \ \sim iid \; \mathcal{N}(0, \sigma^2)$\par }

Note that there is no constant term to capture the intercept. For simplicity of notation we will assume that a constant such as $1$ is included in the vector $x_t$ to allow for a non-zero intercept in the linear expression above. Since we have absorbed the constant $1$ in $x_t$, the corresponding elements of $R_{k0}$ (variance of $\mu_k$) and $M$ (variance of $x_t$ within cluster $k$) should be close to zero to reflect the fact that there is little uncertainty about this element of $x_t$. We represent prior information about the regression vector $\beta_k$ for cluster $k$ through a Gaussian prior distribution with mean $\beta_{k0}$ and variance $Q_{k0}$. 

We will make a simplifying assumption that each observation of input-output data $(x_t,y_{t+1})$ is independent. In case of daily prediction, this assumption implies that each day provides a new independent observation. This is a simplifying assumption as there may be some overlap in market data $x_t$ and $x_{t+1}$ (as they may incorporate data from same prior days). It is possible to to drop the independence assumption and extend the VI framework to the case where $x_t$ is the state of a linear dynamical system as in \cite{pineda}. However, such an approach leads to greater complexity as one must also estimate the parameters that describe the time evolution of the system.

The assumptions described above imply that the observed data (input $x_t$ and output $y_{t+1}$) is generated by the following model:
\begin{center}
\fbox{%
    \parbox{0.8\linewidth}{%
        \textbf{Hyperparameters and prior information}
        \begin{itemize}
            \item $K$ is the number of clusters of input space $x_t$ (where $x_t \in \mathbb{R}^{n}$).
            \item The fraction of times $x_t$ is in cluster $K$ is $\pi_k$. Thus the sum of $\pi_k$ over all $k$ is one.
            \item For cluster $k$, the mean of $x_t$ is $\mu_k$, which is normally distributed with mean $\mu_{k0}$ and covariance $R_{k0}$. In other words, for cluster $k$, $\mu_k \sim \mathcal{N}(\mu_{k0},R_{k0})$. 
            \item If $x_t$ is in cluster $k$ then $x_t$ is normally distributed with  mean $\mu_{k}$ and covariance $M$. In other words if $c_{t(k)}=1$ then $x_t \sim \mathcal{N}(\mu_{k},M)$. Note that while each cluster has a different mean $\mu_k$, we are assuming the the same covariance $M$ for $x_t$ in each cluster.
            \item The regression coefficient for cluster $k$, which is $\beta_k \in \mathbb{R}^{n}$, is normally distributed with mean $\beta_{k0}$ and covariance $Q_{k0}$. In other words, $\beta_k \sim \mathcal{N}(\beta_{k0},Q_{k0})$.
            \item If $x_t$ at time $t$ is in cluster $k$, then the output $y_{t+1}$ that we would like to predict is normally distributed with mean $x_t' \beta_k$ and standard deviation $\sigma$ (equivalently, if $c_{t(k)}=1$, then $y_{t+1}  \ \sim \mathcal{N}(x_t' \beta_k, \sigma)$). 
            \item Summary of hyperparameters: $\pi$ for cluster probabilities; $\sigma$ for standard deviation of exogenous noise in market change; $(\mu_{k0},R_{k0})$ for mean and variance of each cluster mean; $M$ for variance of $x_t$ within a given cluster; $(\beta_{k0},Q_{k0})$ regression parameter mean and its variance for each cluster. 
        \end{itemize}
        \textbf{Generative process}
        \begin{enumerate}
            \item For $\; k \in \left[ 1, \hdots, K \right]$, draw $\mu_k \ \sim \mathcal{N}(\mu_{k0},R_{k0})$
            \item For $k \in \left[ 1, \hdots, K \right]$, draw $\beta_k \ \sim \mathcal{N}(\beta_{k0},Q_{k0})$ 
            \item For $t \in \left[ 1, \hdots T \right]$
            \begin{enumerate}
                \item Draw latent variable $c_{t(k)} \ \sim \ \text{Cat} (\pi_k)$ which describes the cluster assignment for time $t$. $c_{t(k)}$ is an indicator variable which is 1 for only one of its $K$ elements and zero for all other $K-1$ elements.
                \item If the cluster assignment in previous step is $k$, draw input $x_t  \ \sim \mathcal{N}(\mu_{k},M)$
                \item If the cluster assignment is $k$, draw output $y_{t+1}  \ \sim \mathcal{N}(x_t' \beta_k, \sigma^2)$
            \end{enumerate}
        \end{enumerate}
    }
}
\end{center}

Let us illustrate the above generative process with an example. Consider the case when there are $K=3$ clusters for the market data $x_t$. One can think of the three clusters as three market regimes corresponding to market rallies, sell-offs, and relatively unchanged markets (note that the algorithm will identify the clusters and they may not have this intuitive interpretation). At any given time, the markets are in one of these three possible states or clusters. For the three possible states, the input or the market condition as defined by $x_t$ is normally distributed with mean corresponding to that cluster and variance $M$. The  output $y_{t+1}$ for time $t+1$ is assumed to be normally distributed with mean $x_t' \beta_k$ and standard deviation $\sigma$.  

Given data $x_t$ and $y_{t+1}$ for $t \in \left[ 1, \hdots T \right]$, the goal is to obtain posterior distribution for cluster means $\mu_k$, regression vectors $\beta_k$, and cluster assignment variables $c_t$ for all data points indexed by time $t$. 

We now summarize the above prior information where $\pi_k$, $\mu_{k0}$, $R_{k0}$, $\beta_{k0}$ and $Q_{k0}$ for $k \in \{1, \cdots, K\}$ are hyperparameters for cluster assignments, means and regression vectors:
\begin{align}
\label{cluster}
p(c_{t(k)}=1) & = \pi_k \; \; \text{for  k } \in \{1,\cdots,K\} \; \text{where } \sum_1^K \pi_k=1 \\
\label{muprior}
    p(\mu_k ; \mu_{k0},R_{k0}) & =  \frac{\text{exp}(-\frac{1}{2}(\mu_k-\mu_{k0})'R_{k0}^{-1}(\mu_k-\mu_{k0}))}{\sqrt{(2 \pi)^n|R_{k0}|}} \; \; \; \text{for  k } \in \{1,\cdots,K\}\\
\label{betaprior}
     p(\beta_k ;\beta_{k0}, Q_{k0}  ) & =  \frac{\text{exp}(-\frac{1}{2}(\beta_k-\beta_{k0})'Q_{k0}^{-1}(\beta_k-\beta_{k0}))}{\sqrt{(2 \pi)^{n}|Q_{k0}|}} \; \; \; \text{for  k } \in \{1,\cdots,K\}
 \end{align}
Based on the above, $x_t$ and $y_{t}$ are assumed to have the following probability densities where $M$ and $\sigma$ are hyperparameters:
\begin{align}
\label{xprior}
    p(x_t | \mu_k, c_{t(k)}=1) & =  \frac{\text{exp}(-\frac{1}{2}(x_t-\mu_{k})'M^{-1}(x_t-\mu_{k}))}{\sqrt{(2 \pi)^n|M|}} \\
 \label{yprior} 
 p(y_{t+1} | x_t, \beta_k, c_{t(k)}=1) & =  \frac{\text{exp}(-\frac{1}{2 \sigma^2}(y_{t+1}-x_t' \beta_k)^2)}{\sigma \sqrt{(2 \pi)}}
\end{align}
Let us assume we have $T$ observations of the data available. For ease of communication we will use the following notation
\begin{equation}
\text{\bf{c}}=\{c_1, \hdots, c_T \}\ , \ \text{\bf{x}}=\{x_1, \hdots, x_T \}\ , \ \boldsymbol\beta =\{ \beta_1, \hdots, \beta_K \} \ , \ \boldsymbol\mu =\{ \mu_1, \hdots, \mu_K \} \ , \ \text{\bf{y}}=\{y_2, \hdots, y_{T+1} \}
\end{equation}

With the above generative process, 
\begin{equation}
\label{factorization}
    p(\boldsymbol\beta, \boldsymbol\mu , \text{\bf{c}}, \text{\bf{x}}, \text{\bf{y}})= \prod_{k=1}^K p(\beta_k) \prod_{k=1}^K p(\mu_k) \prod_{t=1}^T \left[ p(c_t) p(x_t|\boldsymbol\mu, c_t) p(y_{t+1} | x_t, \boldsymbol\beta, c_t) \right]
\end{equation}
The above factorization implies
\begin{align}
\label{logP}
\text{log }  p(\boldsymbol\beta, \boldsymbol\mu , \text{\bf{c}}, \text{\bf{x}}, \text{\bf{y}})
& = \sum_{k=1}^K \text{log }  p(\beta_k) + \sum_{k=1}^K \text{log }  p(\mu_k) \nonumber \\
& + \sum_{t=1}^T \left[ \text{log }  p(c_t) + \text{log }  p(x_t |\boldsymbol\mu, c_t) + \text{log }  p(y_{t+1} |\boldsymbol\beta, x_t, c_t) \right]
\end{align}

For approximating the posterior probability density of latent variables $p(\boldsymbol\beta, \boldsymbol\mu , \text{\bf{c}} | \text{\bf{x}}, \text{\bf{y}})$ by variational distribution $q(\boldsymbol\beta, \boldsymbol\mu , \text{\bf{c}})$, we will consider the following mean-field distribution where $\beta$, $\mu$ and $c$ are independent and governed by their own variational parameters (with the variational parameters being $\hat{\beta}_k, \hat{Q}_k,\hat{\mu}_k, \hat{R}_k$ and $\phi_{t}$) :
\begin{equation}
\label{meanfield1}
    q(\boldsymbol\beta, \boldsymbol\mu , \text{\bf{c}})= q_\beta (\boldsymbol\beta)q_\mu (\boldsymbol\mu)q_c (\text{\bf{c}}) = \prod_{k=1}^K q_\beta (\beta_k ; \hat{\beta}_k, \hat{Q}_k) \prod_{k=1}^K q_\mu (\mu_k; \hat{\mu}_k, \hat{R}_k) \prod_{t=1}^T  q_c (c_t ; \phi_t)  
\end{equation}
Please note that the superscript " $\hat{ }$ " will denote variational parameters of the corresponding distribution. The mean-field family of distributions above may not contain the true posterior because of the assumption that $\boldsymbol\beta, \boldsymbol\mu$ and $\text{\bf{c}}$ are independent. The independence assumption leads to much more tractable estimation of the the posterior probability density of the latent variables. The distributions $q_c$ in terms of its variational parameter $\phi_{t}$ are assumed to be as follows:

\begin{equation}
   \label{qc1}
q_c(c_{t(k)}=1)  =\phi_{t_k} \; \; \; \text{for  t } \in \{1,\cdots,T\} \;  \text{and for  k } \in \{1,\cdots,K\} 
\end{equation}

Without making any assumptions on the family of distributions for $q_\beta$ and $q_\mu$, it is shown in the Appendix that these are Gaussian distributions with the following form in terms of their variational parameters:   
\begin{align}
\label{qmu1}
q_\mu(\mu_k ; \hat{\mu}_k, \hat{R}_k) & =  \frac{\text{exp}(-\frac{1}{2}(\mu_k-\hat{\mu}_k)'\hat{R}_k^{-1}(\mu_k-\hat{\mu}_k))}{\sqrt{(2 \pi)^n|\hat{R}_{k}|}} \; \; \; \text{for  k } \in \{1,\cdots,K\} \\
\label{qbeta1}
q_\beta(\beta_k ; \hat{\beta}_k, \hat{Q}_k) & =  \frac{\text{exp}(-\frac{1}{2}(\beta_k-\hat{\beta}_k)'\hat{Q}_k^{-1}(\beta_k-\hat{\beta}_k))}{\sqrt{(2 \pi)^n|\hat{Q}_{k}|}} \; \; \; \text{for  k } \in \{1,\cdots,K\} 
\end{align}

The variational posterior distribution $q(\boldsymbol\beta, \boldsymbol\mu , \bf{c})$, which approximates  $p(\boldsymbol\beta, \boldsymbol\mu , \bf{c} | \bf{x}, \bf{y})$ is obtained by maximizing ELBO (\ref{elbo}). The variational parameters $\{\hat{\beta}_k, \hat{Q}_k,\hat{\mu}_k, \hat{R}_k, \phi_{t_k}\}$ have the following interpretation for the posterior distribution: $\phi_{t_k}$ is the probability of the $k'th$ cluster assignment for data $x_t$; parameters $\hat{\mu}_k, \hat{R}_k$ are the mean and variance of the $k'th$ cluster of input $x_t$; and $\hat{\beta}_k, \hat{Q}_k$ are the mean and variance of the 
regression vector for the $k'th$ cluster. Utilizing (\ref{logP}) and (\ref{meanfield1}), ELBO function (\ref{elbo}) can be expressed as follows where the expectation is with respect to variational distribution $q(\boldsymbol\beta, \boldsymbol\mu , \bf{c})$: 
\begin{align}
\label{elbo1}
    \text{ELBO}(\hat{\beta}_k, \hat{R}_k,\hat{\mu}_k, \hat{Q}_k, \phi_{t_k}) & =  \mathbb{E}_q \left[ \text{log }  p(\boldsymbol\beta, \boldsymbol\mu , \text{\bf{c}}, \text{\bf{x}}, \text{\bf{y}}) \right]
    - \mathbb{E}_q \left[ \text{log } q(\boldsymbol\beta, \boldsymbol\mu , \text{\bf{c}}) \right]
    \nonumber \\
   & = \sum_{k=1}^K \mathbb{E}_q \left[ \text{log} \ p(\mu_k) \right] +  \sum_{k=1}^K \mathbb{E}_q \left[ \text{log} \ p(\beta_k) \right] \nonumber \\
    & + \sum_{t=1}^T \Big( \mathbb{E}_q \left[ \text{log} \ p(c_t) \right] + \mathbb{E}_q \left[ \text{log} \ p(x_t |\boldsymbol\mu, c_t) \right] + \mathbb{E}_q \left[ \text{log} \ p(y_{t+1} |\boldsymbol\beta, x_t, c_t) \right] \Big) \nonumber \\
    & - \sum_{t=1}^T \mathbb{E}_q \left[ \text{log} \ q_c(c_t) \right] - \sum_{k=1}^K \mathbb{E}_q \left[ \text{log} \ q_\mu(\mu_k) \right] - \sum_{k=1}^K \mathbb{E}_q \left[ \text{log} \ q_\beta (\beta_k) \right] 
\end{align}

The variational parameters $\{\hat{\beta}_k, \hat{Q}_k,\hat{\mu}_k, \hat{R}_k, \phi_{t_k}\}$ are to be estimated so that the ELBO function described above is maximized.

Recall that the indicator function $c_{t(k)}$ is $1$ if $x_t$ is in cluster $k$ and $0$ otherwise.  Thus
\begin{align}
\label{xprior1}
    p(x_t | \boldsymbol\mu, c_t) & = \prod_{k=1}^K \left[ p(x_t | \mu_k, c_{t(k)}=1) \right]^{c_{t(k)}}\\
    \label{yprior1}
    p(y_{t+1}| x_t, c_t, \boldsymbol\beta) & = \prod_{k=1}^K \left[ p(y_{t+1} | x_t, \beta_k, c_{t(k)}=1) \right]^{c_{t(k)}}
\end{align}
where $p(x_t | \mu_k, c_{t(k)}=1)$ and $p(y_{t+1} | x_t, \beta_k, c_{t(k)=1})$ are described in (\ref{xprior}) and (\ref{yprior}).

The variational parameters that maximize the ELBO function in (\ref{elbo1}) can now be obtained based on (\ref{meanfield}). It is shown in the Appendix that the optimal variational parameters satisfy the following equations:
\begin{align}
\label{phitk}
    \phi_{t_k} & = \frac{r_{t(k)}}{\sum_{k=1}^K r_{t(k)}} \; \text{ for } k \in \{1, \hdots,K\} \; \text{and } t \in \{1, \hdots,T\} \\
    \hat{R}_{k} & =  \left[R_{k0}^{-1}  + M^{-1} \sum_{t=1}^T \phi_{t_k} \right]^{-1} \\
\hat{\mu}_{k} & =  \hat{R}_{k} \left[ R_{k0}^{-1} \mu_{k0} + M^{-1} \sum_{t=1}^T \phi_{t_k} x_t       \right] \\
\hat{Q}_{k} & =  \left[Q_{k0}^{-1}  + \frac{1}{\sigma^2} \sum_{t=1}^T \phi_{t_k} x_t x_t' \right]^{-1} \\
\hat{\beta}_{k} & = \hat{Q}_{k} \left[ Q_{k0}^{-1} \beta_{k0} + \frac{1}{\sigma^2} \sum_{t=1}^T \phi_{t_k} y_{t+1} x_t       \right]
\end{align}
where $r_{t(k)}$ is defined as follows:
\begin{equation}
\label{rtk1}
    r_{t(k)}  :=\text{exp} \left[ \text{log} (\pi_k) + x_t'M^{-1} \hat{\mu}_k - \frac{1}{2} \text{trace}(M^{-1} (\hat{\mu}_k \hat{\mu}_k'+\hat{R}_k))
    +  \frac{1}{\sigma^2} y_{t+1} x_t' \hat{\beta}_k - \frac{1}{2 \sigma^2} \text{trace}(x_tx_t' \ (\hat{\beta}_k \hat{\beta}_k'+\hat{Q}_k)) \right]
\end{equation}

One of the most commonly used algorithms for obtaining optimal variational parameters is Coordinate Ascent Variational Inference (CAVI) (see for example \cite{BleiStats}). Here parameters are updated one at a time, keeping other parameters constant. The process is repeated until the ELBO converges. The CAVI algorithm for estimating parameters is described in Algorithm \ref{cavialg}. The algorithm may however converge to a local maximum and thus running the algorithm with different initial estimates of the variational parameters can improve the approximated model posterior.

\begin{algorithm}
\caption{CAVI algorithm for estimating variational parameters} \label{cavialg}
\renewcommand{\algorithmicrequire}{\textbf{Input:}}
\renewcommand{\algorithmicensure}{\textbf{Output:}}
\begin{algorithmic}
\REQUIRE Data $x_t$ and $y_{t+1}$ for $t \in \{1, \cdots,T\}$ 
\REQUIRE Hyperparameters: Number of clusters $K$; intra-cluster variance M; $\pi_k$, $\mu_{k0}$, $R_{k0}$, $\beta_{k0}$ and $Q_{k0}$ for $k \in \{1, \cdots, K\}$ 
\STATE{}
\STATE{Initialize variational parameters $\hat{\beta}_k, \hat{Q}_k,\hat{\mu}_k, \hat{R}_k$ for all $k \in \{1, \cdots, K\}$ and $\phi_{t_k}$ for all $t \in \{1, \cdots,T\}$ and $k \in \{1, \cdots, K\}$}
\STATE{}
\WHILE{ELBO has not converged} 
\STATE{}
\FOR{each time step $t \in \{1, \hdots, T\}$}
\FOR{$k \in \{1, \hdots, K\}$}
\STATE{$\phi_{tk} \gets \frac{r_{t(k)}}{\sum_{k=1}^K r_{t_k}}$} \ \ \ \ 
\COMMENT{(where $r_{t_k}$ is as defined in (\ref{rtk1}))}
\ENDFOR
\ENDFOR
\STATE{}
\FOR{$k \in \{1, \hdots, K\}$}
\STATE{$\hat{R}_{k} \gets \left[R_{k0}^{-1}  + M^{-1} \sum_{t=1}^T \phi_{t_k} \right]^{-1}$}
\STATE{$\hat{\mu}_{k} \gets \hat{R}_{k} \left[ R_{k0}^{-1} \mu_{k0} + M^{-1} \sum_{t=1}^T \phi_{t_k} x_t       \right]$}
\STATE{$\hat{Q}_{k} \gets \left[Q_{k0}^{-1}  + \frac{1}{\sigma^2} \sum_{t=1}^T \phi_{t_k} x_t x_t' \right]^{-1}$}
\STATE{$\hat{\beta}_{k} \gets \hat{Q}_{k} \left[ Q_{k0}^{-1} \beta_{k0} + \frac{1}{\sigma^2} \sum_{t=1}^T \phi_{t_k} y_{t+1} x_t       \right]$}
\ENDFOR
\STATE{}
\ENDWHILE
\STATE{}
\RETURN $\phi_{t_k},\hat{R}_{k}, \hat{\mu}_{k},\hat{Q}_{k}$ and  $\hat{\beta}_{k}$
\end{algorithmic}
\end{algorithm}

\section{Predictive density and distribution of predicted output}

Our goal is to estimate the probability density of forecast $y_{t+1}$ based on available information up to time $t$. For the following discussion we will denote all the available observations of $x_t$ and $y_t$ up to time $t$ as $\mathcal{D}_t$, i.e.
\[
\mathcal{D}_t:=\{x_1, \cdots, x_t, y_1, \cdots, y_t\}
\]
One key difference in prediction vs. estimation of variational parameters as described in Algorithm \ref{cavialg} is that in the estimation step the entire data of inputs and output is known. For the case of prediction, the output $y_{t+1}$ that we are trying to predict is not known at time $t$. Thus the variational parameter  $\phi_{tk}$ which describes the probability of cluster $k$ at time $t$, cannot be obtained using equations (\ref{phitk}) and (\ref{rtk1}) since the $r_{t(k)}$ expression includes $y_{t+1}$. For prediction we will denote $q_c^{pred}(c_{t(k)}=1)$ as the variational probability of $c_{t(k)}$ based on data $\mathcal{D}_t$. From (\ref{meanfield}) and (\ref{logP}) one obtains the following where $\mathbb{E}_{-c_{t}}$ is expectation with respect to  $\{\mu_1, .., \mu_K \}$, $\{\beta_1, .., \beta_K \}$ and $\{c_i ; i \neq t \}$
\begin{align}
    q_c^{pred}(c_{t(k)}=1; \phi_t) & \propto \text{exp} \Big( \mathbb{E}_{-c_{t}} \left[ \text{log} \ p(\boldsymbol\beta, \boldsymbol\mu , \text{\bf{c}}, \mathcal{D}_t) \right]   \Big) \nonumber \\
    & \propto \text{exp} \Big( \mathbb{E}_{-c_{t}} \left[ \text{log} \ p(c_{t(k)}=1,x_t | \pi, \boldsymbol\mu ) \right]  + \text{const} \Big) \nonumber \\
    & \propto \text{exp} \Big( \mathbb{E} \left[ \text{log} \ p(c_{t(k)}=1 | \pi ) \right] + \mathbb{E}_\mu \left[ \text{log} \ p(x_t | c_{t(k)}=1, \mu_k ) \right] + \text{const} \Big) \nonumber \\
    & \propto \text{exp} \Big( \text{log} (\pi_k) - \frac{1}{2} x_t'M^{-1} x_t + x_t'M^{-1} \ \mathbb{E}_\mu( \mu_k) - \frac{1}{2} \mathbb{E}_\mu(\mu_k' M^{-1} \mu_k) \ + \ \text{const} \Big) \nonumber \\
    \label{qpred1}
    & \propto \text{exp} \Big( \text{log} (\pi_k) + x_t'M^{-1} \hat{\mu}_k - \frac{1}{2} \text{trace}(M^{-1} (\hat{\mu}_k \hat{\mu}_k'+\hat{R}_k))
     + \ \text{const} \Big)
\end{align}
where "const" represents terms that are independent of $k$, $\mu$ and $\beta$. One key difference between the above and the expression for $q_c^*$ in
(\ref{qcluster}) used in obtaining variational parameters is that the expression involving $p(y_{t+1}|x_t, \beta_k, c_t)$ is not there in (\ref{qpred1}) since $y_{t+1}$ is not known at time $t$. Since $\sum_k q_c^{pred}(c_{t(k)}) =1$, from (\ref{qpred1}) one observes
\begin{align}
\label{qpred2}
    q_c^{pred}(c_{t(k)}=1) & = \frac{\text{exp} \Big( \text{log} (\pi_k) + x_t'M^{-1} \hat{\mu}_k - \frac{1}{2} \text{trace}(M^{-1} (\hat{\mu}_k \hat{\mu}_k'+\hat{R}_k))
     \Big)}{\sum_{j=1}^K \text{exp} \Big( \text{log} (\pi_j) + x_t'M^{-1} \hat{\mu}_j - \frac{1}{2} \text{trace}(M^{-1} (\hat{\mu}_j \hat{\mu}_j'+\hat{R}_j))\Big)} 
\end{align}
For predicting the distribution of $y_{t+1}$, we will approximate the posterior distribution of $\beta$, $\mu$ and $c_t$ by their variational approximation. In particular, the density of the estimate of $y_{t+1}$ is obtained as follows:
\begin{eqnarray}
p(y_{t+1}| \mathcal{D}_t) & =  \
\bigintssss_{\bf{c}} \bigintssss_{\boldsymbol\beta } p(y_{t+1}|x_t, \boldsymbol\beta, \bf{c}) p (\boldsymbol\beta, \bf{c} | \mathcal{D}_t) \ d\boldsymbol\beta \ d\bf{c}  \nonumber \\
& \approx \bigintssss_{\bf{c}} \bigintssss_{\boldsymbol\beta }  p(y_{t+1}|x_t, \boldsymbol\beta, \bf{c}) q (\boldsymbol\beta, \bf{c} ) d\boldsymbol\beta \ d\bf{c}  \nonumber \\
\label{posterior}
& = \sum_{k=1}^K \left[ \bigintsss_{ \beta_k}     p(y_{t+1}|x_t, c_{t(k)}=1, \beta_k)   q_\beta(\beta_k) d\beta_k \right] q_c^{pred}(c_{t(k)}=1)\nonumber \\
\label{density}
& = \sum_{k=1}^K   q_c^{pred}(c_{t(k)}=1) \left[ \bigintsss_{ \beta_k}    \frac{\text{exp}(-\frac{1}{2 \sigma^2}(y_{t+1}-x_t' \beta_k)^2)} {\sigma \sqrt{(2 \pi)}}  \frac{\text{exp}(-\frac{1}{2}(\beta_k-\hat{\beta}_k)'\hat{Q}_k^{-1}(\beta_k-\hat{\beta}_k))}{\sqrt{(2 \pi)^n|\hat{Q}_{k}|}}d\beta_k \right]
\end{eqnarray}

In the above we have used (\ref{qbeta1}) for $q(\beta_k)$, (\ref{yprior}) for $p(y_{t+1} | x_t, \beta_k, c_{t(k)}=1)$, and (\ref{qpred2}) for $q_c^{pred}(c_{t(k)}=1)$. Let us define
\begin{equation}
\label{Delta}
   \Delta_k:= \left( \frac{1}{ \sigma^2}x_tx_t'+ \hat{Q}_k^{-1} \right)^{-1}=\hat{Q}_k-\frac{1}{\sigma^2+x_t'\hat{Q}_k x_t} \  \hat{Q}_k x_t x_t' \hat{Q}_k \; \; \; \; \text{for } \ k=1, \cdots,K 
\end{equation}
\begin{equation}
\label{psi}
   \psi_{ky}:= \frac{1}{\sigma^2}y_{t+1}x_t+\hat{Q}_k^{-1}\hat{\beta}_k
   \; \; \; \text{for } \ k=1, \cdots,K 
\end{equation}
\begin{equation}
\label{betabar}
   \bar{\beta}_k:=\beta_k-\Delta_k \psi_{ky} \; \; \; \text{for } \ k=1, \cdots,K  
\end{equation}
where the second equality in (\ref{Delta}) follows from the Matrix Inversion Lemma. Substituting the above definitions in (\ref{density}), we find that the density of the forecast is:
\begin{eqnarray}
p(y_{t+1}| \mathcal{D}_t) & =\frac{1}{\sigma \sqrt{(2 \pi)}} 
 \sum_{k=1}^K  q_c^{pred}(c_{t(k)}=1) \left[ \bigintsss_{ \beta_k}   \ \frac{ 
 \text{exp}(-\frac{y_{t+1}^2}{2 \sigma^2} -\frac{1}{2} \hat{\beta}'_k\hat{Q}_k^{-1}\hat{\beta}_k +\frac{1}{2}\psi_{ky}'\Delta_k \psi_{ky}
 -\frac{1}{2}\bar{\beta_k}'\Delta_k^{-1}\bar{\beta_k})}{\sqrt{(2 \pi)^n|\hat{Q}_{k}|}} \ d\beta_k \right] \nonumber \\
& = \frac{1}{\sigma \sqrt{(2 \pi)}}  
 \sum_{k=1}^K  \left[ q_c^{pred}(c_{t(k)}=1) \text{exp}(-\frac{y_{t+1}^2}{2 \sigma^2} -\frac{1}{2} \hat{\beta}_k' \hat{Q}_k^{-1}\hat{\beta}_k +\frac{1}{2}\psi_{ky}'\Delta_k \psi_{ky}) \sqrt{\frac{|\Delta_{k}|}{|\hat{Q}_{k}|}} \ 
\bigintsss_{ \Bar{\beta_k}}\frac{\text{exp}(-\frac{1}{2}\bar{\beta_k}'\Delta_k^{-1}\bar{\beta_k})}{\sqrt{(2 \pi)^n|\Delta_{k}|}} d\Bar{\beta_k} \right] \nonumber \\
\label{ydensity}
& = \frac{1}{\sigma \sqrt{(2 \pi)}}
  \sum_{k=1}^K  \left[ q_c^{pred}(c_{t(k)}=1) \  \sqrt{\frac{|\Delta_{k}|}{|\hat{Q}_{k}|}} \  \text{exp}(-\frac{y_{t+1}^2}{2 \sigma^2}-\frac{1}{2} \hat{\beta}_k'\hat{Q}_k^{-1}\hat{\beta}_k +\frac{1}{2}\psi_{ky}'\Delta_k \psi_{ky}) \right]
\end{eqnarray}
In obtaining the last expression above we have used the fact that Gaussian distribution density integrates to one, that is
\[
\bigintssss_{ \Bar{\beta_k}}\frac{\text{exp}(-\frac{1}{2}\bar{\beta_k}'\Delta_k^{-1}\bar{\beta_k})}{\sqrt{(2 \pi)^n|\Delta_{k}|}} d\Bar{\beta_k}=1
\]
Utilizing the definitions of $\Delta_k$ and $\psi_{ky}$ in (\ref{Delta}) and (\ref{psi}), one can show the following after some algebraic manipulations:
\[
-\frac{y_{t+1}^2}{2 \sigma^2}-\frac{1}{2} \hat{\beta}_k'\hat{Q}_k^{-1}\hat{\beta}_k +\frac{1}{2}\psi_{ky}'\Delta_k \psi_{ky} = - \ \frac{1}{2(\sigma^2+x_t'\hat{Q}_kx_t)} \ (y_{t+1}-\hat{\beta}_k'x_t)^2
\]
Substituting the above in (\ref{ydensity}), one obtains the following expression for predictive density:
\begin{equation}
\label{ydensity1}
    p(y_{t+1}| \mathcal{D}_t) \ = \  \frac{1}{\sigma \sqrt{(2 \pi)}}
  \sum_{k=1}^K  \left[ q_c^{pred}(c_{t(k)}=1) \  \sqrt{\frac{|\Delta_{k}|}{|\hat{Q}_{k}|}} \  \text{exp} \left(- \ \frac{1}{2(\sigma^2+x_t'\hat{Q}_kx_t)} \ (y_{t+1}-\hat{\beta}_k'x_t)^2 \right) \right] 
\end{equation}

From the predictive density described in (\ref{ydensity1}), one can obtain the expected value of $y_{t+1}$ as well as various confidence levels for the prediction. Note that the expression (\ref{ydensity1}) is similar to a weighted combination of $K$ Gaussian probability densities, one from each cluster. The term in square brackets is the contribution to the density estimate from each cluster, which is weighted proportionally to cluster probability ($q_c^{pred}(c_{t(k)}=1)$). The predictive density within cluster $k$ is proportional to that of a Gaussian with mean $\hat{\beta}_k'x_t$ and variance $(\sigma^2+x_t'\hat{Q}_kx_t)$; this variance increases as $\hat{Q}_k$, the uncertainty in $\hat{\beta}_k$, increases. If a cluster $k$ has large $\hat{Q}_k$ (large uncertainty in $\hat{\beta}_k$), the predictive density contribution from the cluster decreases due to $\sqrt{|\hat{Q}_k|}$ in the denominator and the distribution becomes less sensitive to $y$ due to $(\sigma^2+x_t'\hat{Q}_k x_t)$ in the denominator of the exponent term. 

Below we summarize the algorithm to estimate probability density of forecast $y_{t+1}$ based on available information up to time $t$.

\textbf{Summary of the algorithm to predict output distribution}
\begin{itemize}
\item From $\mathcal{D}_{t-1}:=\{x_1, \cdots, x_{t-1}, y_1, \cdots, y_{t-1}\}$, the available data up to time $t-1$, estimate the variational parameters $\hat{R}_{k}, \hat{\mu}_{k},\hat{Q}_{k}$ and  $\hat{\beta}_{k}$ based on CAVI algorithm \ref{cavialg}.
\item From data $x_t$ at time $t$, obtain the cluster probabilities $q_c^{pred}(c_{t(k)}=1)$ for all $k \in \{1,\cdots,K\}$ using (\ref{qpred2}).
\item Obtain the probability density of $y_{t+1}$ using (\ref{ydensity1}).
\end{itemize}
Once the density of forecast is obtained, one can compute other desired metrics such as expected value of $y_{t+1}$ or its range for a desired confidence level etc. 

\section{Illustrative example of S\&P Index one day change} 

To illustrate the performance of the proposed approach, we consider one-day forecasting of the S\&P equity index. The model is trained based on the most recent $250$ days (approximately one year of data) of closing prices of chosen macro indices. The objective is to predict S\&P change for the next business day. The states used for predicting next day S\&P change are:
\[
    x_t=\begin{bmatrix}
    \text{S\&P index change over the most recent day}  \\ \text{AGG bond ETF change over the most recent five days} \\ \text{Standard deviation of USD and JPY exchange rate changes over the most recent five days}\\
    1 \end{bmatrix}
\]
The constant term $1$ is to account for nonzero intercept. All the inputs were end of day index levels from Bloomberg and the output to be estimated was the subsequent one-day S\&P close-to-close return. For the S\&P index there were no adjustments made for dividends as they contribute only a small fraction of daily changes. VI parameter estimation was done after normalizing all the inputs and outputs. The first two inputs (recent S\&P and AGG changes) and output (next-day S\&P change) were normalized to their $z$ score based on the previous $250$ observations where $z$ score is defined as (current value minus most recent $250$ day average)/standard deviation of variable over the most recent $250$ days. The third input is the difference between the standard deviation of JPY/USD exchange rate over the most recent five days and the standard deviation of JPY/USD exchange rate over the most recent $250$ days.

\begin{figure}
    \begin{subfigure}{0.48\textwidth}
        \includegraphics[width=\textwidth]{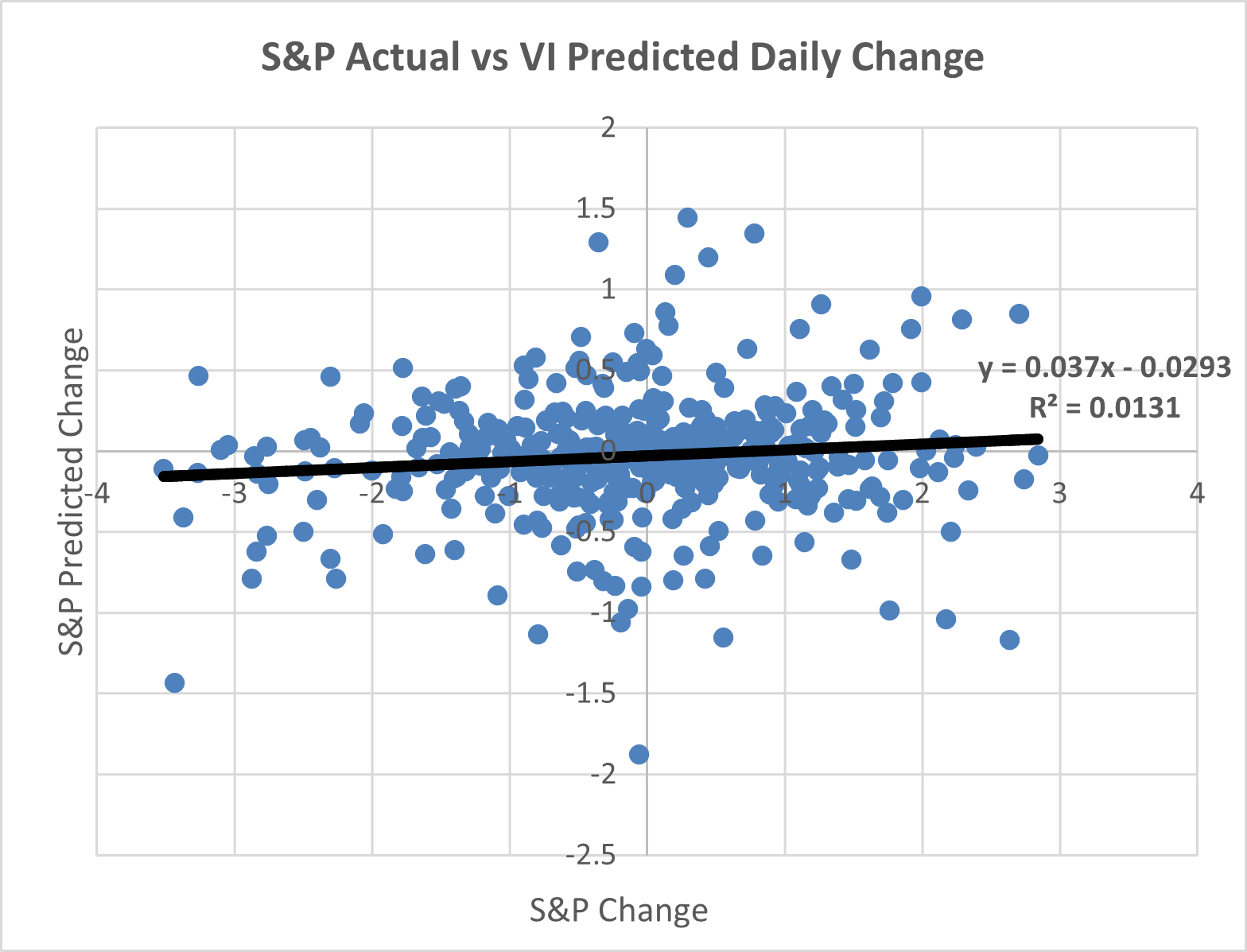}
        \caption{Comparison of normalized daily S\&P changes and VI predicted changes from January 2020 to September 2021.}
    \end{subfigure} \hfill
    \begin{subfigure}{0.48\textwidth}
        \includegraphics[width=\textwidth]{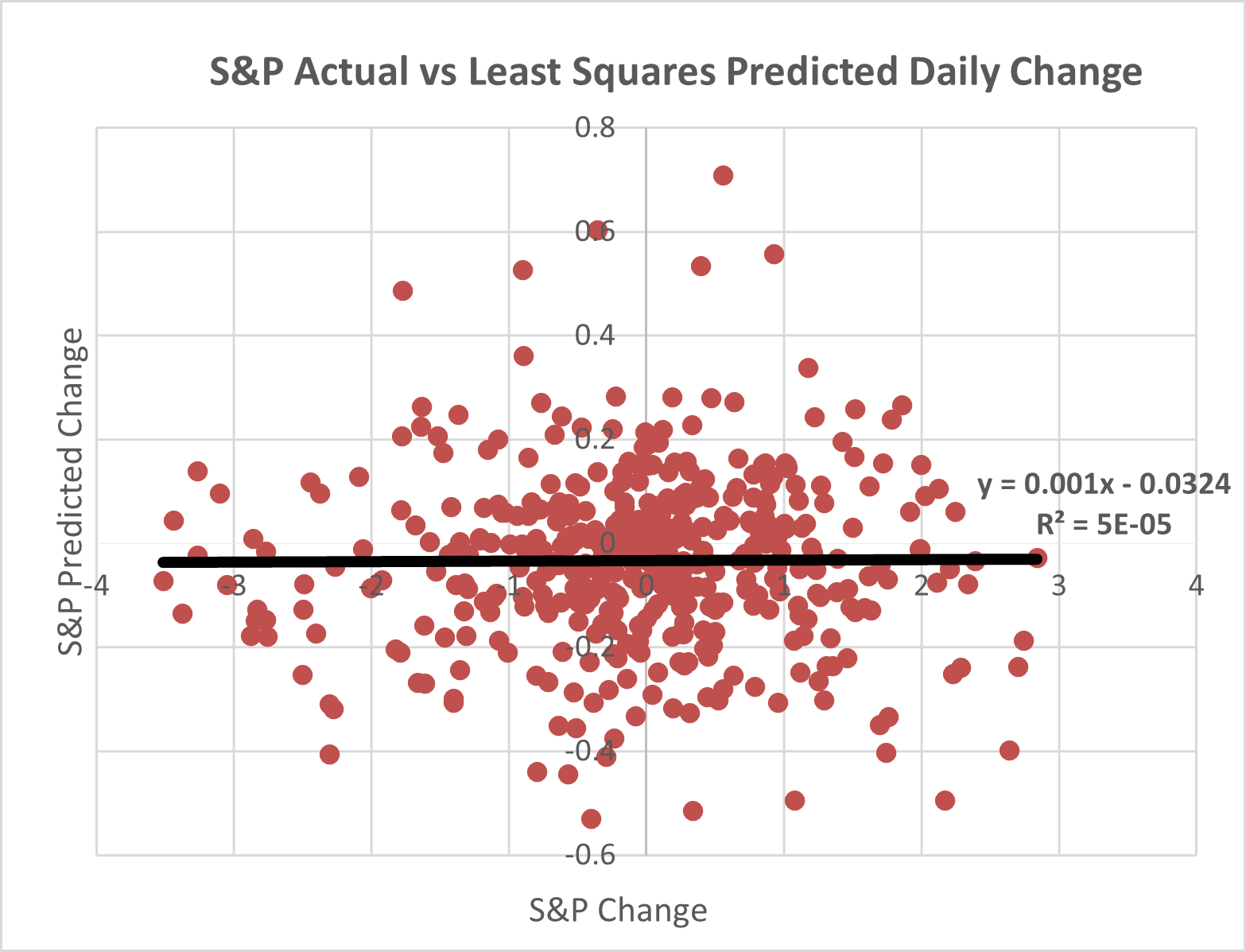}
        \caption{Comparison of normalized daily S\&P and linear regression predicted changes from January 2020 to September 2021.}
    \end{subfigure}
    \caption{Comparison of normalized daily S\&P changes and predicted changes for VI and linear regression utilizing identical data from $250$ previous daily observations. With linear regression (right plot), the slope of the trend line and $R^2$ are almost zero. With VI, $R^2$ is $0.0131$ and the correlation between actual and predicted change is statistically significant at $95\%$ confidence level ($p$-value $=0.017$).}
    \label{exampleSP}
 \end{figure}

For illustrative purposes, only three explanatory variables are included that capture the momentum effect and recent market volatility, but one could include more variables such as those involving customer flows, etc. Figure \ref{exampleSP} provides a scatter plot of $z$ score of actual one-day S\&P change vs. that predicted using the proposed approach with input variable $x_t$ defined above. To show the benefit of cluster based regression over standard regression, a comparison of results based on standard least squares regression using identical data is also provided. The scatter plot covers daily S\&P changes from the beginning of January $2020$ to the end of September $2021$. This period includes both the extreme volatility seen in early $2020$ due to Covid as well as the very benign market conditions and strong market rally seen after June $2020$.  The predicted changes show an observable correlation to the actual changes ($11.5$\% to be precise) that is statistically significant with $95\%$ confidence ($p$-value $=0.017$).

Another way to judge performance is by comparing not the magnitude but the direction of actual S\&P changes to the direction of predicted changes. To address this we analyzed questions such as 1) how often did S\&P increase (respectively, decrease) when the model prediction was for the index to increase (respectively, decrease), and the question with opposite conditioning 2) on the days S\&P increased (respectively, decreased), how often did the forecast predict increase (respectively, decrease)? To make such a comparison we first determined thresholds for actual changes and thresholds for forecasts that would divide the actual and predicted output into three equal buckets for each (each of the three buckets will have equal number of observations). This is done by identifying thresholds $z_{actual}^{low}$ and $z_{actual}^{high}$ such that $\frac{1}{3}$ of actual S\&P changes over the test period were below $z_{actual}^{low}$ and $\frac{1}{3}$ of changes were above $z_{actual}^{high}$. Similarly we identified thresholds $z_{pred}^{low}$ and $z_{pred}^{high}$ such that $\frac{1}{3}$ of predicted S\&P changes over the test period were below $z_{pred}^{low}$ and $\frac{1}{3}$ of changes were above $z_{pred}^{high}$. Note that these thresholds would be different for actual changes compared to predicted changes ($z_{actual}^{low}$ maybe different than $z_{pred}^{low}$). If the model performance is good then one would expect to see a large overlap of observations in the increase (and also decrease) bucket. Table \ref{bucketing} summarizes the predicted vs. realized S\&P changes for the nine combinations of actual change and model predicted change. For example, the first row implies that when the next day prediction was for the S\&P to decrease, $39.04\%$ of the time it decreased, $30.82\%$ of the time it was relatively unchanged, and $30.14\%$ of the time it increased. Similarly, the third row implies that when the prediction was for S\&P to increase, $39.04\%$ of the time it increased, $31.51\%$ of the time it decreased, and $29.45\%$ of the time it was relatively unchanged. If predictions and actual changes were uncorrelated and the three outcomes were equally likely, each element in Table \ref{bucketing} would be close to $33\%$. In Table \ref{bucketing} one observes that all diagonal elements are more than $39\%$, implying that actual changes were broadly correct $39\%$ of the time.   

\begin{table}
\begin{center}
\begin{tabular}{|c|c|c|c|} 
 \hline
 & S\&P Decreased & S\&P Unchanged & S\&P Increased  \\ 
  \hline
Prediction that S\&P will decrease & 39.04\% & 30.82 \%& 30.14 \% \\
  & & & \\
Prediction that S\&P will remain relatively unchanged & 29.25\% & 40.14\% & 30.61 \% \\
& & & \\
Prediction that S\&P will increase & 31.51\% & 29.45\% & 39.04\% \\
 \hline
\end{tabular}
\end{center}
\caption{Daily S\&P changes vs. those predicted from January 2020 to September 2021 where both actual changes and predictions are categorized so that there are equal numbers of observations in each category. If predictions and actual changes were uncorrelated, each term in the table above would be close to $33.33\%$ since each of the three outcomes is equally likely. On the other hand in the case of near-perfect correlations, diagonal elements should be close to $100\%$ and off-diagonal elements close to $0\%$. Notice that the diagonal elements of the matrix exceed $39$\% and are higher than all off-diagonal elements, implying that the actual change is most likely to be in the same category as the prediction.}
\label{bucketing}
\end{table}

Another advantage of the proposed approach is the interpretability of the market regimes/clusters in terms of explanatory variables and the identification of the most important variables for forecasts in each market regime. The variational parameters of the approximate posterior distribution vary with time as they are obtained based on the most recent $250$ days of observations. The results in Table \ref{calmcluster} describe the cluster means (variable $\hat{\mu}_k$) for September 30, 2022.

\begin{table}
\begin{center}
\begin{tabular}{|l |c|c|c| } 
 \hline
Variable & Cluster 1 ($\hat{\mu}_1$) & Cluster 2 ($\hat{\mu}_2$) & Cluster 3 ($\hat{\mu}_3$)  \\ 
  \hline
S\&P 1 day change & 0.0929 & -0.7229 & 0.4672 \\
AGG 5 day change & -0.1759 & -1.2496 & 0.544 \\
JPY/USD 5 day st dev & 0.1948 & 0.0897 & 0.2712 \\
Constant & 1 & 1 & 1 \\
 \hline
\end{tabular}
\end{center}
\caption{Explanatory variable cluster centers for September 30, 2022}
\label{calmcluster}
\end{table}

\begin{table}
\begin{center}
\begin{tabular}{|c|c|c| } 
 \hline
Regression vector cluster 1 ($\hat{\beta}_1$) & Regression vector cluster 2 ($\hat{\beta}_2$) & Regression vector cluster 3 ($\hat{\beta}_3$)  \\ 
  \hline
-0.09 & 0.29 & -0.29 \\
0.20 & -0.02 & 0 \\
-0.62 & 0.73 & -0.06 \\
-1.48 & 0.25 & 0.82 \\
 \hline
\end{tabular}
\end{center}
\caption{Regression parameter estimates for the clusters on September 30, 2022}
\label{calmbeta}
\end{table}

Based on the above tables, one can draw some qualitative conclusions about the clusters which, though relatively simple, may provide useful insight to practitioners. For example, based on the above tables, one may draw the following qualitative conclusions about each of the clusters in terms of a) market conditions, b) important market variables for forecasting in such market conditions, and c) bias for expected change for the next day:

\underline{\it{Cluster 1}}
\begin{itemize}
    \item Market conditions: Based on the first column of Table \ref{calmcluster}, the first cluster corresponds to market conditions where the S\&P is relatively unchanged day-over-day, AGG bond index values are marginally lower over the past five days, and JPY/USD exchange rate volatility over the past five days has exceeded its 250-day average. 
    \item Critical variables for forecast: According to Table \ref{calmbeta}, the most important variable for the forecast is the recent JPY/USD exchange rate volatility (the largest-magnitude element of regression vector $\hat{\beta}_1$ is $-0.62$, corresponding to the $5$-day volatility of the JPY/USD exchange rate relative to its 250-day average). Increasing JPY/USD volatility in such market conditions implies lower S\&P forecasts for the next day because of the negative sign of this parameter.
    \item Bias for the next day S\&P change: The last element of regression vector $\hat{\beta}_1$, corresponding to the intercept, is $-1.48$ (a negative number) and thus the expected S\&P change for the next day is negative in cluster 1 assuming smaller contributions from other variables.
\end{itemize}

\underline{\it{Cluster 2}}
\begin{itemize}
    \item Market conditions: The mean values of this cluster represent conditions where the S\&P has decreased day-over-day and bond prices have declined significantly over the previous five days (observations based on the second column of Table \ref{calmcluster}).    
    \item Critical variables for forecast: The two important regression parameters are the coefficients related to S\&P change ($0.29$) and JPY/USD volatility ($0.73$). The S\&P change regression parameter can be interpreted as a momentum effect - the next day change based on this factor is proportional to $0.29$ times the previous day change. From the JPY/USD volatility regression parameter of $0.73$, one concludes that elevated JPY/USD five-day volatility relative to its 250-day average implies higher S\&P forecasts for the next day. The regression parameter for AGG is quite small ($-0.02$), which implies that AGG change does not have much impact on predicted S\&P change in this cluster.  
    \item Bias for the next day S\&P change: Assuming that the contributions from other variables are small, S\&P changes are expected to be small as the last element of regression vector $\hat{\beta}_2$, corresponding to the intercept, is $0.25$,  which is relatively small. 
\end{itemize}

\underline{\it{Cluster 3}}
\begin{itemize}
    \item Market conditions: In this cluster, S\&P and AGG values have increased and JPY/USD volatility over the past five days is higher than its long-term average (conclusion based on the third column of Table \ref{calmcluster}).    
    \item Critical variables for forecast: The regression parameter corresponding to previous day S\&P change is $-0.29$. This implies that based on this factor, the next day predicted change is negatively correlated to previous day S\&P change. Regression parameters for AGG change and JPY/USD volatility are small, implying that their values are not very important for predicting the next-day change. 
    \item Bias for the next day S\&P change: Since the last element of regression vector $\hat{\beta}_3$, corresponding to the intercept, is positive ($0.82$), the S\&P is likely to increase the next day. 
\end{itemize}
Note that predictions have different sensitivities to market variables for different clusters. For example, in clusters 2 and 3, regression parameters related to S\&P change have opposite sign ($0.29$ and $-0.29$ respectively). Additionally, it is important to note that variational parameters will change depending on the historical data used in parameter estimation and thus will likely change from day to day if the estimation is performed based on the most recent historical data. 

\section{Summary}
In this paper, an approach based on variational inference is proposed for simultaneously identifying clusters and cluster-specific regression parameters that provide good prediction of output. In the proposed approach, clusters of explanatory variables are identified such that the predicted output is a linear function of the selected inputs, where the linear function depends on the cluster. One advantage of the proposed approach in financial applications is that it identifies market regimes in which future changes share a similar dependence on input variables. It is shown that a fairly simple model with only three explanatory variables provides useful predictions of financial indices. Another advantage of the proposed approach is that it also provides the predictive density of the forecast. Due to the broad applicability of the problem formulation and the computational efficiency of the proposed algorithm, the approach can be useful in a wide range of applications, not limited to the financial domain.

\section*{Appendix}

We now derive the parameters for the variational distribution based on (\ref{meanfield}). First let us consider the variational density of the cluster mean $\mu_k$. From (\ref{meanfield})
\[
 q_\mu^*(\mu_k; \hat{\mu}_{k}, \hat{R}_{k}) \; \propto \text{exp} \Big( \mathbb{E}_{-\mu_k} \left[ \text{log} \ p(\boldsymbol\beta, \boldsymbol\mu , \text{\bf{c}}, \text{\bf{x}}, \text{\bf{y}}) \right] \Big)
\]
where $\mathbb{E}_{-\mu_k}$ is expectation with respect to $\{\beta_1, \cdots, \beta_K\}$, $\{c_i \ , \ 1 \leq i \leq T \}$ and $\{\mu_i \ , \ i \neq k.  \}$. From (\ref{muprior}), (\ref{xprior}), (\ref{xprior1}),  and variational parameters for $q_c$ in (\ref{qc1}) one notes that
\begin{align}
    \label{qmu}
    q_\mu^*(\mu_k; \hat{\mu}_{k}, \hat{R}_{k}) & \propto \text{exp} \Big( \mathbb{E}_{-\mu_k} \left[ \text{log} \ p(\boldsymbol\beta, \boldsymbol\mu , \text{\bf{c}}, \text{\bf{x}}, \text{\bf{y}}) \right] \Big)  \nonumber \\
    & \propto \text{exp} \Big(  \text{log} \ p(\mu_k ; \mu_{k0}, R_{k0} ) +  \sum_{t=1}^T  \mathbb{E}_{c_t} \left[ \text{log} \ p(x_t | c_{t(k)}, \mu_k ) \right] \Big) + \text{const}  \nonumber \\
    & \propto \text{exp} \Big(  -\frac{1}{2}(\mu_k-\mu_{k0})'R_{k0}^{-1}(\mu_k-\mu_{k0}) 
    + \sum_{t=1}^T \mathbb{E}_{c_t} \left[ c_{t(k)}=1 \right] \ \text{log} \ p(x_t | \mu_k ) \Big) + \text{const} \Big) \nonumber \\
    & \propto \text{exp} \Big( \mu_k'R_{k0}^{-1}\mu_{k0} - \frac{1}{2} \mu_k'R_{k0}^{-1}\mu_k
    + \sum_{t=1}^T \phi_{tk} \left[ -\frac{1}{2}(x_t-\mu_{k})'M^{-1}(x_t-\mu_{k}) \right]+ \text{const} \Big)
    \nonumber \\
    & \propto \text{exp} \Big( - \frac{1}{2} (\mu_k -\hat{\mu}_{k})' \hat{R}_{k}^{-1}(\mu_k -\hat{\mu}_{k}) + 
     \text{const} \Big)
\end{align}
where "const" includes constant terms that do not depend on $\mu_k$, $\mathbb{E}_{c_t}$ is expectation with respect to $q_c$ and the variational parameters $\hat{\mu}_{k}$ and $\hat{R}_{k}$ are defined as follows:
\begin{align}
    \label{hatmuR}
    \hat{R}_{k} & := \left[R_{k0}^{-1}  + M^{-1} \sum_{t=1}^T \phi_{tk} \right]^{-1} \nonumber \\
    \hat{\mu}_{k} & := \hat{R}_{k} \left[ R_{k0}^{-1} \mu_{k0} + M^{-1} \sum_{t=1}^T \phi_{tk} x_t       \right]
\end{align}
From the exponential distribution in (\ref{qmu}) one notes that the variational distribution of $\mu_k$, described by $q_\mu^*(\mu_k)$, is Gaussian with mean $\hat{\mu}_{k}$ and variance $\hat{R}_{k}$. Note that in obtaining the above, we did not make any assumptions about the distribution family of $q_\mu^*(\mu_k)$.

We next consider the variational density of $\beta_k$. From (\ref{betaprior}), (\ref{yprior1}) and variational parameters for $q_c$ in (\ref{qc1}) one notes that
\begin{align}
    \label{qbeta}
    q_\beta^*(\beta_k; \hat{\beta}_{k}, \hat{Q}_{k}) & \propto \text{exp} \Big( \mathbb{E}_{-\beta_k} \left[ \text{log} \ p(\boldsymbol\beta, \boldsymbol\mu , \text{\bf{c}}, \text{\bf{x}}, \text{\bf{y}}) \right] \Big)  \nonumber \\
    & \propto \text{exp} \Big(  \text{log} \ p(\beta_k ; \beta_{k0}, Q_{k0} ) +  \sum_{t=1}^T  \mathbb{E}_{c_t} \left[ \text{log} \ p(y_{t+1} | c_{t(k)}, x_t, \beta_k ) \right] \Big) + \text{const}  \nonumber \\
    & \propto \text{exp} \Big(  -\frac{1}{2}(\beta_k-\beta_{k0})'Q_{k0}^{-1}(\beta_k-\beta_{k0}) 
    + \sum_{t=1}^T \mathbb{E}_{c_t} \left[ c_{t(k)}=1 \right] \ \text{log} \ p(y_{t+1} | x_t, \beta_k ) \Big) + \text{const} \Big) \nonumber \\
    & \propto \text{exp} \Big( \beta_k'Q_{k0}^{-1}\beta_{k0} - \frac{1}{2} \beta_k'Q_{k0}^{-1}\beta_k
    + \sum_{t=1}^T \phi_{tk} [ -\frac{1}{2 \sigma^2}(y_{t+1}-\beta_{k}'x_t)^2 ]+ \text{const} \Big)
    \nonumber \\
    & \propto \text{exp} \Big( \beta_k'Q_{k0}^{-1}\beta_{k0} - \frac{1}{2} \beta_k'Q_{k0}^{-1}\beta_k
    + \sum_{t=1}^T \phi_{tk} [ \frac{1}{\sigma^2}y_{t+1}\beta_{k}'x_t -  \frac{1}{2 \sigma^2} \beta_{k}'x_tx_t'\beta_k] + \text{const} \Big)
    \nonumber \\
    & \propto \text{exp} \Big(  - \frac{1}{2} (\beta_k -\hat{\beta}_{k})' \hat{Q}_{k}^{-1}(\beta_k -\hat{\beta}_{k}) + 
     \text{const} \Big)
\end{align}
where "const" includes terms that do not depend on $\beta_k$, and the variational parameters $\hat{\beta}_{k}$ and $\hat{Q}_{k}$ are defined as follows:
\begin{align}
    \label{hatbetaQ}
    \hat{Q}_{k} & := \left[Q_{k0}^{-1}  + \frac{1}{\sigma^2} \sum_{t=1}^T \phi_{tk} x_t x_t' \right]^{-1} \nonumber \\
    \hat{\beta}_{k} & := \hat{Q}_{k} \left[ Q_{k0}^{-1} \beta_{k0} + \frac{1}{\sigma^2} \sum_{t=1}^T \phi_{tk} y_{t+1} x_t       \right]
\end{align}

From the exponential distribution in (\ref{qbeta}) one notes that the variational distribution of $\beta_k$, described by $q_\beta^*(\beta_k)$, is Gaussian with mean $\hat{\beta}_{k}$ and variance $\hat{Q}_{k}$. Note that in obtaining the above, we did not make any assumptions about the distribution family of $q_\beta^*(\beta_k)$.

We finally consider the probability for cluster assignment $k$ at time $t$. From (\ref{meanfield}) and (\ref{logP}) one notes that
\begin{align}
    \label{qcluster}
    q_c^*(c_{t(k)}=1; \phi_t) & \propto \text{exp} \Big( \mathbb{E}_{-c_{t}} \left[ \text{log} \ p(\boldsymbol\beta, \boldsymbol\mu , \text{\bf{c}}, \text{\bf{x}}, \text{\bf{y}}) \right] \Big) \ \ \ ( \text{expectation with respect to all random variables other than } c_t ) \nonumber \\
    & \propto \text{exp} \Big( \mathbb{E}_{-c_{t}} \left[ \text{log} \ p(c_{t(k)}=1,x_t,y_{t+1} | \pi, \boldsymbol\mu_, \boldsymbol\beta ) \right]  + \text{const} \Big) \nonumber \\
    & \propto \text{exp} \Big( \mathbb{E} \left[ \text{log} \ p(c_{t(k)}=1 | \pi ) \right] + \mathbb{E}_\mu \left[ \text{log} \ p(x_t | c_{t(k)}=1, \mu_k ) \right] + 
    \mathbb{E}_\beta \left[ \text{log} \ p(y_{t+1} | x_t , \beta_k, c_{t(k)}=1 ) \right] + \text{const} \Big)
\end{align}
where $\mathbb{E}_{-c_{t}}$ is expectation with respect to $\{\beta_1, \cdots, \beta_K\}$, $\{\mu_1, \cdots, \mu_K \}$ and $\{c_i ; i \neq t \}$ and the expectation of all the terms that do not depend on $c_t(k)$ are lumped into the constant term. $\mathbb{E}_\mu$ and $\mathbb{E}_\beta$ denote expectations with respect to variational distributions $q_\mu(\mu)$ and $q_\beta(\beta)$ respectively. The first term is the log prior of $c_{t(k)}$:  
\begin{equation}
    \label{ElogC}
    \mathbb{E} \left[ \text{log} \ p(c_{t(k)}=1 | \pi ) \right] = \text{log} (\pi_k) \ \; \; \ \text{(from (\ref{cluster}))}
\end{equation}
From (\ref{xprior}) and (\ref{xprior1}) the expected log probability of data $x_t$ is:
\begin{align}
\label{Elogx}
\mathbb{E}_{\mu} \left[ \text{log} \ p(x_t | c_{t(k)}=1 ) \right] &  
  =  \mathbb{E}_\mu  \text{log} \ p(x_t | \mu_k ) \nonumber \\
& =  -\frac{n}{2} \text{log} 2 \pi- \frac{1}{2} \text{log} |M| - \frac{1}{2} x_t'M^{-1} x_t + x_t'M^{-1} \ \mathbb{E}_\mu( \mu_k) - \frac{1}{2} \text{trace}(M^{-1} \ \mathbb{E}_\mu (\mu_k \mu_k'))  \nonumber \\
& =   x_t'M^{-1} \ \mathbb{E}_\mu ( \mu_k) - \frac{1}{2} \text{trace}(M^{-1} \ \mathbb{E}_\mu (\mu_k \mu_k'))  + \text{const} 
\nonumber \\
& =  x_t'M^{-1} \hat{\mu}_k - \frac{1}{2} \text{trace}(M^{-1} (\hat{\mu}_k \hat{\mu}_k'+\hat{R}_k))  + \text{const}
\end{align}
where "const" denotes a constant term that does not depend on $c_{t(k)}$. In obtaining the last step above we have used the fact under the variational distribution $q^*_\mu$ described in (\ref{qmu}), $\mu_k$ is normally distributed with mean $\hat{\mu}_k$ and variance $\hat{R}_k$.

From (\ref{yprior}) and (\ref{yprior1}), the expected log probability of data $y_{t+1}$, which is the last term in (\ref{qcluster}), is:

\begin{align}
\label{Elogy}
\mathbb{E}_{\beta} \left[ \text{log} \ p(y_{t+1}| c_{t(k)}=1, x_t ) \right] & = 
  \text{log} \ p(y_{t+1}| c_{t}, x_t, \beta_k ) \nonumber \\
& = -\frac{1}{2} \text{log} 2 \sigma^2 \pi - \frac{1}{ 2\sigma^2} y_{t+1}^2 + \frac{1}{ \sigma^2} y_{t+1} x_t' \ \mathbb{E}_\beta( \beta_k) - \frac{1}{ 2\sigma^2} \text{trace}(x_tx_t' \ \mathbb{E}_\beta (\beta_k \beta_k'))  \nonumber \\
& =   \frac{1}{\sigma^2} y_{t+1} x_t' \hat{\beta}_k - \frac{1}{2\sigma^2} \text{trace}(x_tx_t' \ (\hat{\beta}_k \hat{\beta}_k'+\hat{Q}_k))   + \text{const} 
\end{align}
where as before "const" is a term that does not depend on $c_{t(k)}$. In obtaining the last step above we have used the fact under the variational distribution $q^*_\beta$ described in (\ref{qbeta}), $\beta_k$ is normally distributed with mean $\hat{\beta}_k$ and variance $\hat{Q}_k$.

Combining (\ref{qcluster}), (\ref{ElogC}), (\ref{Elogx}), and (\ref{Elogy}), the variational update for the $k'th$ cluster assignment at time $t$ is
\begin{equation}
    \label{clustervar}
    q_c^*(c_{t(k)}; \phi_t) \propto  r_{t(k)} 
\end{equation}
where 
\begin{equation}
    \label{rtk}
    r_{t(k)}:=\text{exp} \left[ \text{log} (\pi_k) + x_t'M^{-1} \hat{\mu}_k - \frac{1}{2} \text{trace}(M^{-1} (\hat{\mu}_k \hat{\mu}_k'+\hat{R}_k))
    +  \frac{1}{\sigma^2} y_{t+1} x_t' \hat{\beta}_k - \frac{1}{2 \sigma^2} \text{trace}(x_tx_t' \ (\hat{\beta}_k \hat{\beta}_k'+\hat{Q}_k)) \right]
\end{equation}
Since the sum of probabilities of cluster assignments is one, from (\ref{clustervar}) and (\ref{rtk})
\begin{equation}
    \label{clusterprob}
    q_c^*(c_{t(k)}=1)=\phi_{tk} = \frac{r_{t(k)}}{\sum_{k=1}^K r_{t(k)}}
\end{equation}

\bibliographystyle{unsrt}  

\begin{thebibliography}{1}

\bibitem{Bishop}
C. M. Bishop. 
\newblock Pattern Recognition and Machine Learning.
\newblock  Springer-Verlag, 2006.

\bibitem{BleiStats}
D. M. Blei, A. Kucukelbir and J. D. McAuliffe. 
\newblock Variational Inference : A Review for Statisticians.
\newblock {\em arXiv.1601.00670}, 2018.

\bibitem{dele}
A. Deleforge, F. Forbes, and R. Horaud. 
\newblock High-dimensional regerssion with gaussian mixtures and partially-latent response variables.
\newblock {\em Statistics and Computing}, vol. 25, no. 5, 2015.

\bibitem{drugo}
J. Drugowitsch. 
\newblock Variational Bayesian inference for linear and logistic regression.
\newblock {\em arXiv.1310.5438}, 2013.

\bibitem{Enke}
D. Enke and S. Thawornwong.
\newblock The use of data mining and neural networks for forecasting stock market returns.
\newblock {\em Expert Systems with Applications}, Volume 29, Issue 4, November 2005.

\bibitem{Gelfand}
A. Gelfand and A. Smith. 
\newblock Sampling based approaches to calculating marginal densities.
\newblock {\em Journal of the American Statistical Association}, vol 85, pp 398-409, 1990.

\bibitem{hassan}
M. R. Hassan and B. Nath.
\newblock Stock Market Forecasting Using Hidden Markov Model: A New Approach.
\newblock {\em Proceedings of the 2005 5th International Conference on Intelligent Systems Design and Applications (ISDA’05)}, 2005.

\bibitem{hoffman}
M. D. Hoffman, D. M. Blei, C. Wang and J. Paisley.
\newblock Stochastic Variational Inference.
\newblock {\em Journal of Machine Learning Research}, vol 14, 2013.

\bibitem{Kim}
K. Kim.
\newblock Financial time series forecasting using support
vector machines.
\newblock {\em 	Neurocomputing}, vol. 55, 2003.

\bibitem{Khaidem}
L. Khaidem, S. Saha and S. R. Dey.
\newblock Predicting the direction of stock market prices using random forest.
\newblock {\em 	arXiv:1605.00003}, 2016.

\bibitem{pineda}
X. Alameda-Pineda, V. Drouard and R. Horaud. 
\newblock Variational inference and learning of piecewise-linear dynamical systems.
\newblock {\em arXi:.2006.01668}, 2020.

\bibitem{shen}
J. Shen, M. O. Shafiq. 
\newblock Short-term stock market price trend prediction using a comprehensive deep learning system. 
\newblock {\em J Big Data}, 7, 66 (2020). https://doi.org/10.1186/s40537-020-00333-6

\bibitem{shridhar}
K. Shridhar, F. Laumann nd M. Liwicki. 
\newblock A Comprehensive guide to Bayesian Convolutional Neural
Network with Variational Inference. 
\newblock {\em https://arxiv.org/abs/1901.02731v1.}

\bibitem{tan}
S. L. Tan and D. J. Nott. 
\newblock Variational approximation for mixtures of linear mixed models.
\newblock {\em Journal of Computational and Graphical Statistics}, vol. 23, no. 2, 2014.


\end{thebibliography}

\end{document}